\documentclass[aps,pre,twocolumn,superscriptaddress,showpacs]{revtex4}
\usepackage{epsfig,amsmath,graphicx,amssymb}
\usepackage[colorlinks=false,linkcolor=blue]{hyperref}
\usepackage[header,title,page,titletoc]{appendix}
\usepackage{graphicx}

\usepackage{color}

\def\be{\begin{equation}}
\def\ee{\end{equation}}
\def\bea{\begin{eqnarray}}
\def\eea{\end{eqnarray}}

\begin{document}
\title{Magnetic deformation theory of a vesicle}
\author{Yao-Gen Shu}
\affiliation{CAS Key Laboratory of Theoretical Physics, Institute of Theoretical Physics, Chinese Academy of Sciences, Beijing, China}
\author{Zhong-Can Ou-Yang}
\email{oy@mail.itp.ac.cn}
\affiliation{CAS Key Laboratory of Theoretical Physics, Institute of Theoretical Physics, Chinese Academy of Sciences, Beijing, China}

\date{\today}

\begin{abstract}
We have extended the Helfrich's spontaneous curvature model [M. Iwamoto and Z. C. Ou-Yang. Chem. Phys. Lett. \textbf{590}(2013)183; Y. X. Deng, et.al., EPL. \textbf{123}(2018)68002] of the equilibrium vesicle shapes by adding the interaction between magnetic field and the constituent molecules to explain the phenomena of the reversibly deformation of artificial stomatocyte[P. G. van Rhee, et.al., Nat. Commun. \textbf{Sep 24;5:5010}(2014)doi: 10.1038/ncomms6010.] and the anharmonic deformation of a self-assembled nanocapsules of bola-amphiphilic molecules and the linear birefringence[O.V. Manyuhina, et.al., Phys. Rev. Lett. \textbf{98}(2007)146101.]. However, the sophistic mathematics in differential geometry is still covered. Here, we present the derivations of formulas in detailed to reveal the perturbation of deformation $\psi$ under two cases.
\end{abstract}


\maketitle


\section{instruction}
The spontaneous curvature model\cite{helfrich73} of the equilibrium shapes and deformations
of lipid bilayer vesicles, which has been proposed by Helfrich for
more than four decades, was used to successfully explain the
biconcave discoid shape of red blood cells\cite{helfrich76,ouyang93,ouyang961}, so that it is well accepted in biophysics\cite{nossal91}. Particularly, it predicted that the anchor rings generates circles of radii in the ratio of $1/\sqrt{2}$\cite{ouyang90}, and the ratio was precisely confirmed by experiments in toroidal vesicles\cite{mutz91}, phospholipid membrane\cite{rudolph91} and micelles\cite{lin94}. Recently, the curvature elasticity model has been
extended to investigate shapes in soft matter, such as the helical
structures in carbon nanotubes\cite{ouyang97} and in bile ribbons\cite{ouyang98}, cylindrical
structures in the smectic-A phase\cite{naito93} and in peptide nanotubes\cite{ouyang08}, the circle-domain instability in lipid monolayers\cite{iwamoto04} and icosahedral structures in virus capsids\cite{ouyang10,ouyang101}.

If a vesicle is assembled from diamagnetic amphiphilic block-copolymers with a highly anisotropic magnetic susceptibility, we can manipulate its deformation by an external magnetic field. For example, the artificial stomatocyte\cite{Rhee14} can be reversibly opened and closed by varying an external magnetic field. The artificial stomatocyte, thus, has a great potential to transport drug to a target. On the other hand, the small deformation can be measured by birefringence\cite{man07}. The magnetic deformation theory had been proposed by adding the interaction between the magnetic field and the constituent molecules into the shape energy\cite{iwamoto13,ouyang18}, the experimental data were explained satisfactorily.

However, the sophistic mathematics in differential geometry is still covered. Here, we present the derivations of formulas in detailed to reveal the perturbation of deformation $\psi$ under two cases.

\section{free energy of a vesicle in magnetic field}\label{s:1}
Physically, the shape of the vesicle is finally determined by the equilibrium state, at which the energy of any physical system must be at its minimum, i.e. the equilibrium energy of a vesicle must be less than that of other deformation induced by a slightly perturbation. Helfrich proposed that the shape energy of a vesicle can be given by
\begin{eqnarray}
F_1&\equiv&\frac{1}{2}\kappa_{\rm c}\oint(2H+c_0)^2{\rm d}A+\Delta p\int {\rm d}V+\lambda \oint {\rm d}A, \label{eq:1}
\end{eqnarray}
where $\kappa_{\rm c}$ is the bend modulus of vesicle membrane, $H\equiv-(c_1+c_2)/2$ is the mean value of the two principal curvatures($c_1, c_2$), $c_0$ is the spontaneous curvature, $\Delta p\equiv p_{\rm out}-p_{\rm in}$ is the difference pressure of transmembrane, $\lambda$ is the tensile stress acting on the membrane. Mathematically, $\Delta p$ and $\lambda$ may be considered as Lagrange multipliers.

If the vesicle is assembled from diamagnetic amphiphilic block-copolymers with a highly anisotropic magnetic susceptibility and is in a magnetic field, the interaction between the magnetic field ($\vec{\boldsymbol{\mathcal{H}}}$)
and the constituent molecules($F_{\rm B}\equiv-\frac{1}{2}\Delta \chi t\mu\oint(\vec{\boldsymbol{\mathcal{H}}}\cdot\vec{\boldsymbol{n}})^2{\rm d}A$) has to be added into the shape energy.
\begin{eqnarray}
F&\equiv&F_1+F_{\rm B}\nonumber\\
&=&\frac{\kappa_{\rm c}}{2}\oint \left(2H+c_0\right)^2{\rm d}A+\Delta p\int {\rm d}V+\lambda \oint {\rm d}A\nonumber\\
&& -\frac{1}{2}\Delta \chi t\mu\oint\left(\vec{\boldsymbol{\mathcal{H}}}\cdot\vec{\boldsymbol{n}}\right)^2{\rm d}A\label{eq:2}
\end{eqnarray}
where $t$ is the thickness of the membrane of vesicle, $\vec{\boldsymbol{n}}$ is the outward unit normal and $\Delta\chi\equiv\chi_\shortparallel-\chi_{\perp}$, in which $\chi$ is the diamagnetic susceptibility, while $\chi_\shortparallel$ and $\chi_{\perp}$ are diamagnetic susceptibility parallel and perpendicular to $\vec{\boldsymbol{n}}$ respectively.

\section{shape equation of a vesicle}
In order to find the shape equation of the vesicle, it is necessary to calculate the first variation of $F$. For small deformation
\begin{eqnarray}
\vec{\boldsymbol{r}}=\vec{\boldsymbol{r}}_0+\psi\vec{\boldsymbol{n}},\label{eq:3}
\end{eqnarray}
we get\cite{naito95,tu2017}
\begin{eqnarray}
&&\delta^{(1)} F_1=\Delta p\int\delta^{(1)} ({\rm d}V)+\lambda \oint\delta^{(1)}({\rm d}A)\nonumber\\
&&+\frac{\kappa_{\rm c}}{2}\oint \left[\left(2H+c_0\right)^2\delta^{(1)} ({\rm d}A)+4\left(2H+c_0\right)\delta^{(1)} H{\rm d}A\right]\nonumber\\
&=&\oint\left[\Delta p-2\lambda H+\kappa_{\rm c}\left(2H+c_0\right)\left(2H^2-c_0H-2K\right)\right.\nonumber\\
&&\left.+2\kappa_{\rm c}\nabla ^2H\right]\psi \sqrt{g}{\rm d}u{\rm d}v,\label{eq:4}
\end{eqnarray}
in which we used the relations\cite{oy1989}
\begin{eqnarray}
\delta^{(1)} ({\rm d}V)&=&\psi\sqrt{g}{\rm d}u{\rm d}v,\nonumber\\
\delta^{(1)} ({\rm d}A)&=&-2H\psi\sqrt{g}{\rm d}u{\rm d}v,\nonumber\\
\delta^{(1)} H&=&(2H^2-K+\frac{1}{2}\nabla^2)\psi,\nonumber
\end{eqnarray}
and $K\equiv c_1c_2$ is a Gaussian curvature, $\nabla^2c_0=0$ due to $c_0$ is a constant.

The variation of $F_{\rm B}$ for the situation of uniform $\vec{\boldsymbol{\mathcal{H}}}$ ($\delta\vec{\boldsymbol{\mathcal{H}}}=0$) is
\begin{eqnarray}
\delta^{(1)} F_{\rm B}&=&-\frac{t\Delta\chi\mu}{2}\delta^{(1)}\left(\oint(\vec{\boldsymbol{\mathcal{H}}}\cdot\vec{\boldsymbol{n}})^2{\rm d}A\right)\nonumber\\
&=&-\frac{t\Delta\chi\mu}{2}\oint\left[\left(\vec{\boldsymbol{\mathcal{H}}}\cdot\vec{\boldsymbol{n}}\right)^2\delta^{(1)}({\rm d}A)\right.\nonumber\\
&&\left.+2\left(\vec{\boldsymbol{\mathcal{H}}}\cdot\vec{\boldsymbol{n}}\right)\left(\vec{\boldsymbol{\mathcal{H}}}\cdot\delta^{(1)}\vec{\boldsymbol{n}}\right){\rm d}A\right]\nonumber\\
&=&-\frac{t\Delta\chi\mu}{2}\oint\left[\left(\vec{\boldsymbol{\mathcal{H}}}\cdot\vec{\boldsymbol{n}}\right)^2\left(-2H\psi\right)\right.\nonumber\\ &+&\left.2\left(\vec{\boldsymbol{\mathcal{H}}}\cdot\vec{\boldsymbol{n}}\right)\left(\vec{\boldsymbol{\mathcal{H}}}\cdot\left(-g^{kl}\psi_l\vec{\boldsymbol{r}}_k\right)\right)\right]\sqrt{g}{\rm d}u{\rm d}v\label{eq:5}
\end{eqnarray}
where $\delta^{(1)}\vec{\boldsymbol{n}}=-g^{kl}\psi_l\vec{\boldsymbol{r}}_k$\cite{naito95,tu2017}.

To calculate $2\left(\vec{\boldsymbol{\mathcal{H}}}\cdot\vec{\boldsymbol{n}}\right)\left(\vec{\boldsymbol{\mathcal{H}}}\cdot\left(-g^{kl}\psi_l\vec{\boldsymbol{r}}_k\right)\right)$, we set $\vec{\boldsymbol{U}}\equiv-2\left(\vec{\boldsymbol{\mathcal{H}}}\cdot\vec{\boldsymbol{n}}\right)\vec{\boldsymbol{\mathcal{H}}}$, then
\begin{eqnarray}
&&2\left(\vec{\boldsymbol{\mathcal{H}}}\cdot\vec{\boldsymbol{n}}\right)\left(\vec{\boldsymbol{\mathcal{H}}}\cdot\left(-g^{kl}\psi_l\vec{\boldsymbol{r}}_k\right)\right)=\vec{\boldsymbol{U}}\cdot\left(g^{kl}\vec{\boldsymbol{r}}_k\psi_l\right)\nonumber\\
&&=\vec{\boldsymbol{U}}\cdot\nabla^\prime\psi=\nabla^\prime\cdot\left(\psi\vec{\boldsymbol{U}}\right)-\psi\nabla^\prime\cdot\vec{\boldsymbol{U}}\label{eq:6}
\end{eqnarray}
Furthermore, we assume $\vec{\boldsymbol{W}}\equiv\psi\vec{\boldsymbol{U}}=P\vec{\boldsymbol{u}}+Q\vec{\boldsymbol{v}}+R\vec{\boldsymbol{n}}$, then
\begin{eqnarray}
\oint\nabla^\prime\cdot\vec{\boldsymbol{W}}{\rm d}A&=&\oint\frac{1}{\sqrt{g}}\partial_i\left(\sqrt{g}W^i\right){\rm d}A+\oint\nabla^\prime\cdot\left(R\vec{\boldsymbol{n}}\right){\rm d}A\nonumber\\
&=&\oint\frac{1}{\sqrt{g}}\partial_i\left(\sqrt{g}W^i\right)\sqrt{g}{\rm d}u{\rm d}v\nonumber\\
&&+\oint\left(R\nabla^\prime\cdot\vec{\boldsymbol{n}}+\vec{\boldsymbol{n}}\cdot\nabla^\prime R\right){\rm d}A\nonumber\\
&=&\oint\left(\partial_u\left(\sqrt{g}P\right)+\partial_v\left(\sqrt{g}Q\right)\right){\rm d}u{\rm d}v\nonumber\\
&&+\oint\left(R\nabla^\prime\cdot\vec{\boldsymbol{n}}\right){\rm d}A\nonumber\\
&=&\oint\sqrt{g}P{\rm d}v+\oint\sqrt{g}Q{\rm d}u+\oint R\left(-2H\right){\rm d}A\nonumber\\
&=&-\oint 2HR{\rm d}A\nonumber\\
&=&-\oint 2H\left(\vec{\boldsymbol{W}}\cdot\vec{\boldsymbol{n}}\right){\rm d}A\label{eq:7}
\end{eqnarray}
According to (\ref{eq:6}), $\oint 2\left(\vec{\boldsymbol{\mathcal{H}}}\cdot\vec{\boldsymbol{n}}\right)\left(\vec{\boldsymbol{\mathcal{H}}}\cdot\left(-g^{kl}\psi_l\vec{\boldsymbol{r}}_k\right)\right)\sqrt{g}{\rm d}u{\rm d}v=\oint\left(\vec{\boldsymbol{U}}\cdot\nabla^\prime\psi\right)\sqrt{g}{\rm d}u{\rm d}v$, the second part in square brackets of Eq.(\ref{eq:5}) can be simplified into:
\begin{eqnarray}
&&\oint\left(\vec{\boldsymbol{U}}\cdot\nabla^\prime\psi\right)\sqrt{g}{\rm d}u{\rm d}v\nonumber\\
&=&\oint\left(\nabla^\prime\cdot\left(\psi\vec{\boldsymbol{U}}\right)-\psi\nabla^\prime\cdot\vec{\boldsymbol{U}}\right)\sqrt{g}{\rm d}u{\rm d}v\nonumber\\
&=&\oint \left(-2H\psi\vec{\boldsymbol{U}}\cdot\vec{\boldsymbol{n}}-\psi\nabla^\prime\cdot\vec{\boldsymbol{U}}\right)\sqrt{g}{\rm d}u{\rm d}v\nonumber\\
&=&\oint\left[-2H\psi\left(-2\left(\vec{\boldsymbol{\mathcal{H}}}\cdot\vec{\boldsymbol{n}}\right)\vec{\boldsymbol{\mathcal{H}}}\right)\cdot\vec{\boldsymbol{n}}\right.\nonumber\\
&&\left.-\psi\nabla^\prime\cdot\left(-2\left(\vec{\boldsymbol{\mathcal{H}}}\cdot\vec{\boldsymbol{n}}\right)\vec{\boldsymbol{\mathcal{H}}}\right)\right]\sqrt{g}{\rm d}u{\rm d}v\nonumber\\
&=&\oint\left[4H\left(\vec{\boldsymbol{\mathcal{H}}}\cdot\vec{\boldsymbol{n}}\right)^2\right.\nonumber\\
&&\left.+2\nabla^\prime\cdot\left(\left(\vec{\boldsymbol{\mathcal{H}}}\cdot\vec{\boldsymbol{n}}\right)\vec{\boldsymbol{\mathcal{H}}}\right)\right]\psi\sqrt{g}{\rm d}u{\rm d}v \label{eq:8}
\end{eqnarray}
So the Eq.(\ref{eq:5}) becomes:
\begin{eqnarray}
\delta^{(1)} F_{\rm B}&=&-\frac{t\Delta\chi\mu}{2}\oint\left[\left(\vec{\boldsymbol{\mathcal{H}}}\cdot\vec{\boldsymbol{n}}\right)^2\left(-2H\right) +4H\left(\vec{\boldsymbol{\mathcal{H}}}\cdot\vec{\boldsymbol{n}}\right)^2\right.\nonumber\\
&&\left.+2\nabla^\prime\cdot\left(\left(\vec{\boldsymbol{\mathcal{H}}}\cdot\vec{\boldsymbol{n}}\right)\vec{\boldsymbol{\mathcal{H}}}\right)\right]\psi
\sqrt{g}{\rm d}u{\rm d}v\nonumber\\
&=&-t\Delta\chi\mu\oint\left[H\left(\vec{\boldsymbol{\mathcal{H}}}\cdot\vec{\boldsymbol{n}}\right)^2\right.\nonumber\\
&&\left.+\nabla^\prime\cdot\left(\left(\vec{\boldsymbol{\mathcal{H}}}\cdot\vec{\boldsymbol{n}}\right)\vec{\boldsymbol{\mathcal{H}}}\right)\right]\psi\sqrt{g}{\rm d}u{\rm d}v\label{eq:9}
\end{eqnarray}
Combining Eq.(\ref{eq:4}) and Eq.(\ref{eq:9}), as well as $\delta^{(1)} F=\delta^{(1)} F_1+\delta^{(1)} F_{\rm B}=0$, then
\begin{eqnarray}
\Delta p-2\lambda H+\kappa_{\rm c}\left(2H+c_0\right)\left(2H^2-c_0H-2K\right)+2\kappa_{\rm c}\nabla ^2H\nonumber
\end{eqnarray}
\begin{eqnarray}
=t\Delta\chi\mu\left[H\left(\vec{\boldsymbol{\mathcal{H}}}\cdot\vec{\boldsymbol{n}}\right)^2+\nabla^\prime\cdot\left(\left(\vec{\boldsymbol{\mathcal{H}}}\cdot\vec{\boldsymbol{n}}\right)\vec{\boldsymbol{\mathcal{H}}}\right)\right]\label{eq:10}
\end{eqnarray}
This is an important formula. It gives mathematically the condition that the energy of the vesicle has an extreme value.

If the vesicle is a sphere and there is no magnetic interaction, Eq.(\ref{eq:10}) will directly lead to:
\begin{eqnarray}
\Delta pr_0^3+2\lambda r_0^2-\kappa_{\rm c}c_0r_0\left(2-c_0r_0\right)=0\label{eq:11}
\end{eqnarray}

For a stable vesicle, it is necessary that $\delta^{(2)} F$ is positively definite for any $\psi\ne0$, which has been discussed in Ref\cite{tu2017} in detailed for the situation of $\vec{\boldsymbol{\mathcal{H}}}=0$.

\section{calculation of small deformation}

For $\vec{\boldsymbol{\mathcal{H}}}\ne0$, shape equation Eq.(\ref{eq:10}) is too complex to be analyzed. For perturbation $\psi$ in Eq.(\ref{eq:3}), however, we can variate the left of Eq.(\ref{eq:10}) to find the $\psi$ in spherical coordinates
\begin{equation}
\vec{\boldsymbol{r}}_0=r_0\left(\cos\phi \sin\theta,\sin\phi \sin\theta,\cos\theta\right), \label{eq:12}
\end{equation}
where, $r_0$ is determined by Eq.(\ref{eq:11}). We define\cite{tu2017,tu2004}
\begin{eqnarray}
\bar{\nabla}^2&=&\frac{1}{\sqrt{g}}\partial_i\left(KL^{ij}\sqrt{g}\partial_j\right)\nonumber\\
\bar{\bar{\nabla}}^2&=&\frac{1}{\sqrt{g}}\partial_i\left(Hg^{ij}\sqrt{g}\partial_j\right)\nonumber
\end{eqnarray}
then
\begin{eqnarray}
\delta H&=&\left(2H^2_0-K_0\right)\psi+\frac{1}{2}g^{ij}\nabla_i\psi_j\nonumber\\
&=&\left(2H^2_0-K_0+\frac{1}{2}\nabla^2\right)\psi\label{eq:13}\\
\delta K&=&\left(2H_0K_0+\bar{\nabla}^2\right)\psi\label{eq:14}\\
\delta(\nabla^2H)&=&(\delta\nabla^2)H_0+\nabla^2(\delta H)\label{eq:142}
\end{eqnarray}

For convenient follow-up analysis, we designate the left of Eq.(\ref{eq:10}) as \textbf{left}; while the right of that as \textbf{right}.
\subsection{The variation of the \textbf{left}}
We can variate \textbf{left} to find the perturbation $\psi$ due to \textbf{right} $\ne0$:
\begin{eqnarray}
\delta({\rm \textbf{left}})&=&-2\delta\lambda H_0-2\lambda\delta H\nonumber\\
&&+\kappa_{\rm c}\left(2\delta H\right)\left(2H_0^2-c_0H_0-2K_0\right)\nonumber\\
&&+\kappa_{\rm c}\left(2H_0+c_0\right)\left(4H_0\delta H-c_0\delta H-2\delta K\right)\nonumber\\
&&+2\kappa_{\rm c}\left(\delta\nabla^2\right)H_0+2\kappa_{\rm c}\nabla^2\delta H\label{eq:15}
\end{eqnarray}
with the operates:
\begin{eqnarray}
\nabla^2&=&\dfrac{1}{\sqrt{g}}\partial_i\left(g^{ij}\sqrt{g}\partial_j\right)\nonumber\\
\left(\delta\nabla^2\right)&=&-\dfrac{1}{2}\dfrac{\delta g}{g^{3/2}}\partial_i\left(g^{ij}\sqrt{g}\partial_j\right)\nonumber\\
&&+\dfrac{1}{\sqrt{g}}\partial_i\left(\delta g^{ij}\sqrt{g}\partial_j+g^{ij}\dfrac{\delta g}{2\sqrt{g}}\partial_j\right)\nonumber\\
&=&\dfrac{2H_0}{g}\psi\partial_i\left(g^{ij}\sqrt{g}\partial_j\right)\nonumber\\
&&+\dfrac{1}{\sqrt{g}}\partial_i\left[\left(2\left(2H_0g^{ij}-K_0L^{ij}\right)\psi\sqrt{g}\right.\right.\nonumber\\
&&\left.\left.-2g^{ij}H_0\sqrt{g}\psi\right)\partial_j\right]\nonumber\\
&=&2H_0\psi\nabla^2\nonumber\\
&&+\dfrac{2}{\sqrt{g}}\partial_i\left[\left(H_0g^{ij}-K_0L^{ij}\right)\psi\sqrt{g}\partial_j\right]\nonumber\\
\left(\delta\nabla^2\right)H_0&=&2H_0\psi\nabla^2H_0+2\psi\left(\bar{\bar{\nabla}}^2-\bar{\nabla}^2\right)H_0\nonumber\\
&&+2\left(H_0g^{ij}-K_0L^{ij}\right)\psi_iH_j\nonumber
\end{eqnarray}
and spherical vesicle conditions (see Appendix), the Eq.(\ref{eq:15}) becomes
\begin{eqnarray}
\delta({\rm \textbf{left}})&=&-2\delta\lambda H_0+\left[-2\lambda+\kappa_{\rm c}\left(12H^2_0-4K_0-c_0^2\right)\right]\nonumber\\
&&\left(2H^2_0-K_0+\frac{1}{2}\nabla^2\right)\psi\nonumber\\
&&-2\kappa_{\rm c}\left(2H_0+c_0\right)\left(2H_0K_0+\bar{\nabla}^2\right)\psi\nonumber\\
&&+4\kappa_{\rm c}\left[\psi\left(H_0\nabla^2+\bar{\bar{\nabla}}^2-\bar{\nabla}^2\right)H_0\right.\nonumber\\
&&\left.+\left(H_0g^{ij}-K_0L^{ij}\right)\psi_iH_j\right]\nonumber\\
&&+2\kappa_{\rm c}\nabla^2\left[\left(2H^2_0-K_0+\frac{1}{2}\nabla^2\right)\psi\right]\label{eq:17}
\end{eqnarray}
Considering $\bar{\nabla}^2=\frac{1}{\sqrt{g}}\partial_i\left(KL^{ij}\sqrt{g}\partial_j\right)=-\frac{1}{r_0}\nabla^2$, we can further simplify the Eq (\ref{eq:17}) into:
\begin{eqnarray}
\delta({\rm \textbf{left}})&=&\frac{2}{r_0}\delta\lambda+\left[-2\lambda+\kappa_{\rm c}\left(\frac{8}{r_0^2}-c_0^2\right)\right]\left(\frac{1}{r_0^2}+\frac{1}{2}\nabla^2\right)\psi\nonumber\\
&&-2\kappa_{\rm c}\left(-\frac{2}{r_0}+c_0\right)\left(-\frac{2}{r_0^3}+\bar{\nabla}^2\right)\psi\nonumber\\
&&+2\kappa_{\rm c}\nabla^2\left[\left(\frac{1}{r_0^2}+\frac{1}{2}\nabla^2\right)\psi\right]\label{eq:18}
\end{eqnarray}
\subsection{The \textbf{right}}
The right side of Eq.(\ref{eq:10}) is
\begin{eqnarray}
{\rm \textbf{right}}&=&t\Delta\chi\mu\left[H_0\left(\vec{\boldsymbol{\mathcal{H}}}\cdot\vec{\boldsymbol{n}}\right)^2+\nabla\cdot\left(\vec{\boldsymbol{\mathcal{H}}}\left(\vec{\boldsymbol{\mathcal{H}}}\cdot\vec{\boldsymbol{n}}\right)\right)\right]\nonumber\\
&=&\frac{t\Delta\chi B^2}{\mu}\left[-\frac{\cos^2\theta}{r_0}-\frac{\sin\theta}{r_0}\partial_\theta\cos\theta\right]\nonumber\\
&=&q_1(1-2\cos^2\theta)\nonumber\\
&\equiv& g_0Y_0+g_2Y_2\label{eq:19}
\end{eqnarray}
where $q_1\equiv t\Delta\chi B^2/(\mu r_0)$, $Y_l$ is a spherical harmonics, and
\begin{eqnarray}
g_0&=&\frac{q_1\sqrt{4\pi}}{3},\nonumber\\
g_2&=&-\frac{4q_1}{3}\sqrt{\frac{4\pi}{5}}\nonumber
\end{eqnarray}
On the other hand, Eq.(\ref{eq:18}) also can be calculated if we set $\psi=\sum a_lY_l$ :
\begin{eqnarray}
&&\delta({\rm \textbf{left}})=\frac{2}{r_0}\delta\lambda+\sum\Bigg[ \left(-2\lambda+\kappa_{\rm c}\left(\frac{8}{r_0^2}-c_0^2\right)\right)\nonumber\\
&&\left(\frac{1}{r_0^2}-\frac{l(l+1)}{2r_0^2}\right)-2\kappa_{\rm c}\left(\frac{2}{r_0}-c_0\right)\left(\frac{2}{r_0^3}-\frac{l(l+1)}{r_0^3}\right)\nonumber\\
&&+2\kappa_{\rm c}\frac{-l(l+1)}{r_0^2}\left(\frac{1}{r_0^2}-\frac{l(l+1)}{2r_0^2}\right)\Bigg]a_lY_l\nonumber\\
&=&\frac{2}{r_0}\delta\lambda-\sum\frac{\kappa_{\rm c}}{r_0^4}\left[\frac{2\lambda r_0^2}{\kappa_{\rm c}}+c_0^2r_0^2-4c_0r_0+2l(l+1)\right]\nonumber\\
&&\left[1-\frac{l(l+1)}{2}\right]a_lY_l\label{eq:47}
\end{eqnarray}

\section{results}
Comparing Eq.(\ref{eq:19}) with Eq.(\ref{eq:47}), we can select $\psi=a_0Y_0+a_2Y_2$ according to $\delta(\textbf{left})=\textbf{right}$. Here, we consider two cases: one is the birefringence\cite{helfrich732}, which can be directly used to measure the small deformation of a vesicle, another is the reversible open and closing of a stomatocyte. The two cases lead to the same deformation, an ellipsoid.

\begin{figure}[!t]
\centering
\includegraphics[width=7.92cm]{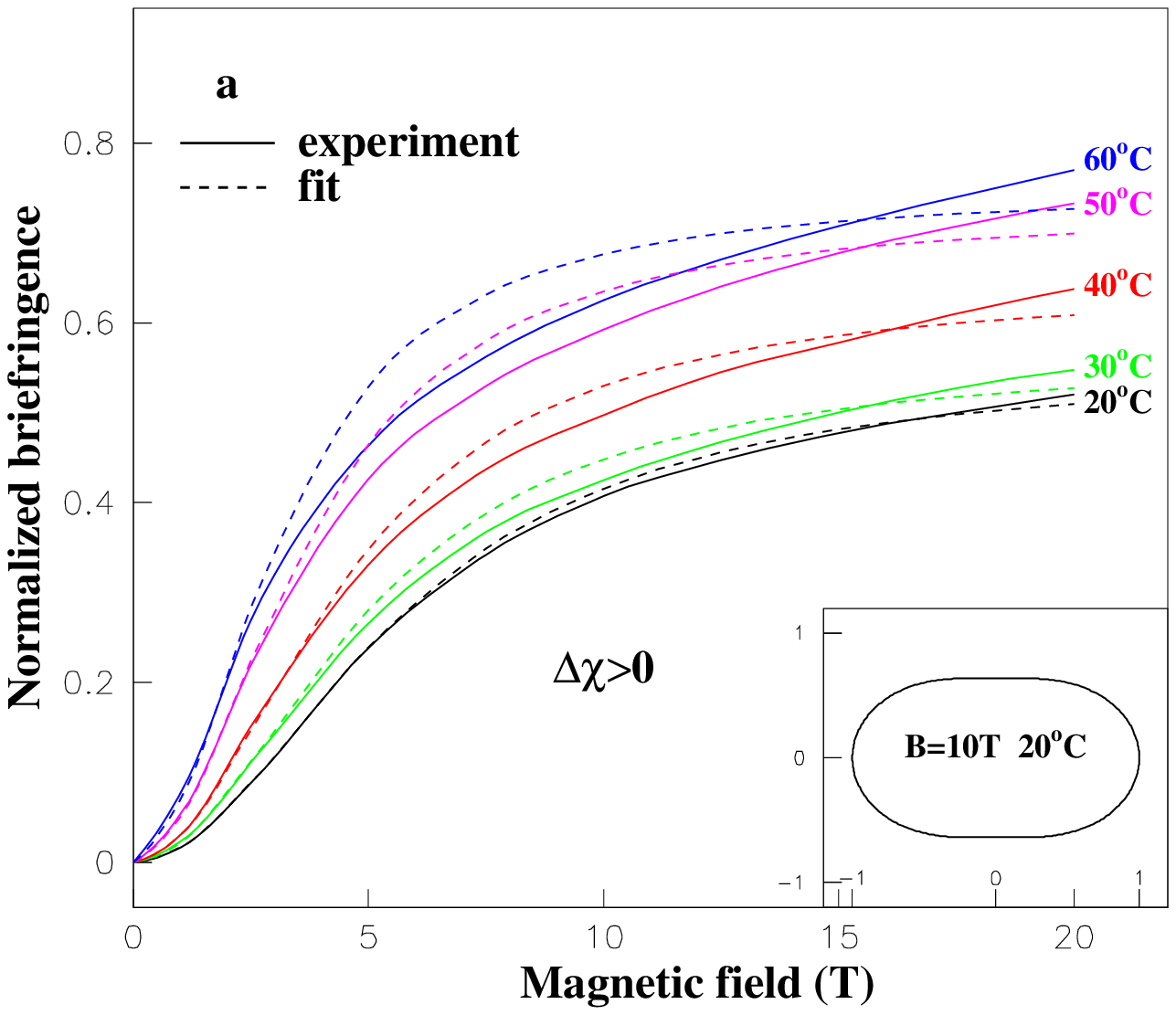}
\includegraphics[width=7.92cm]{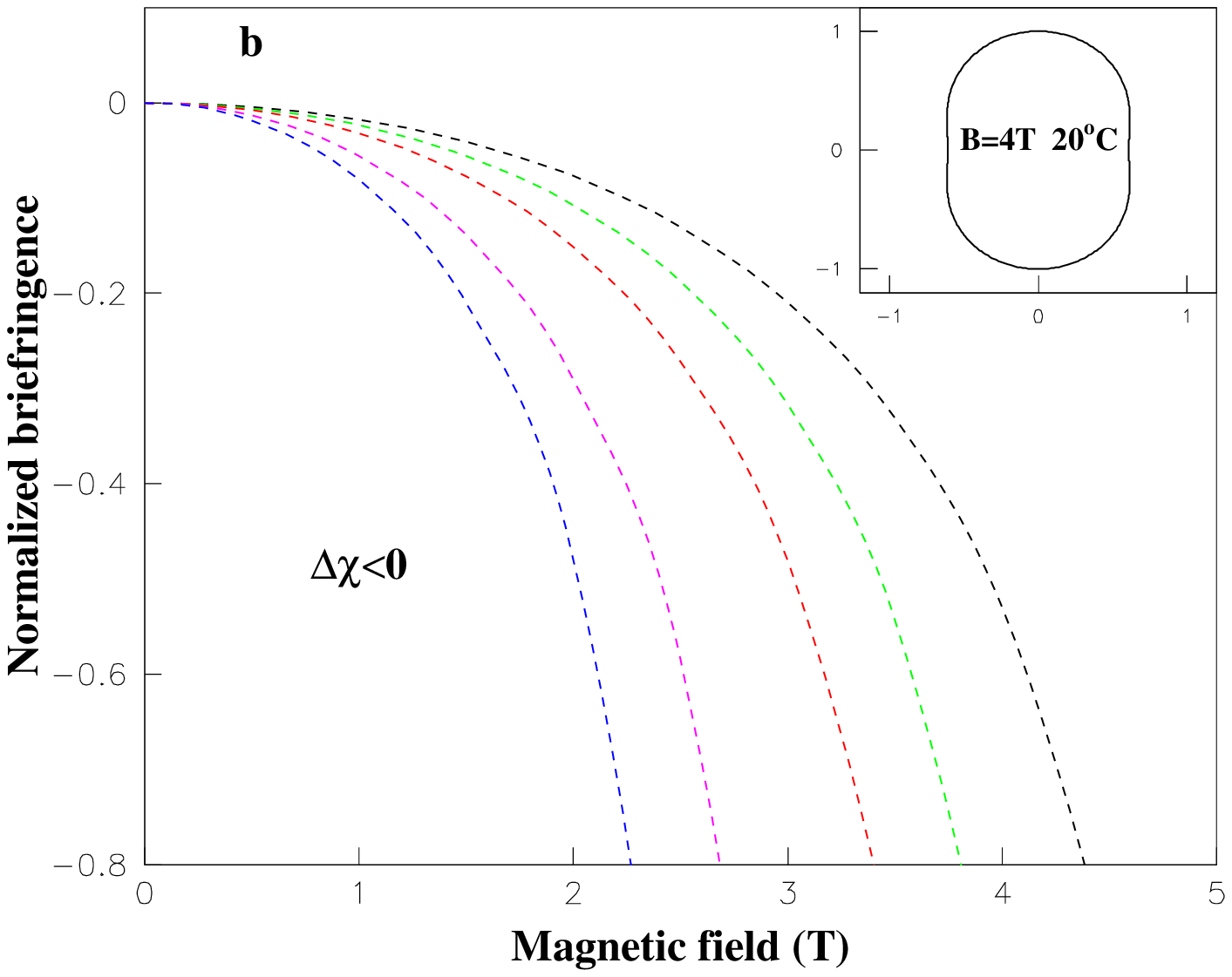}
\includegraphics[width=7.93cm]{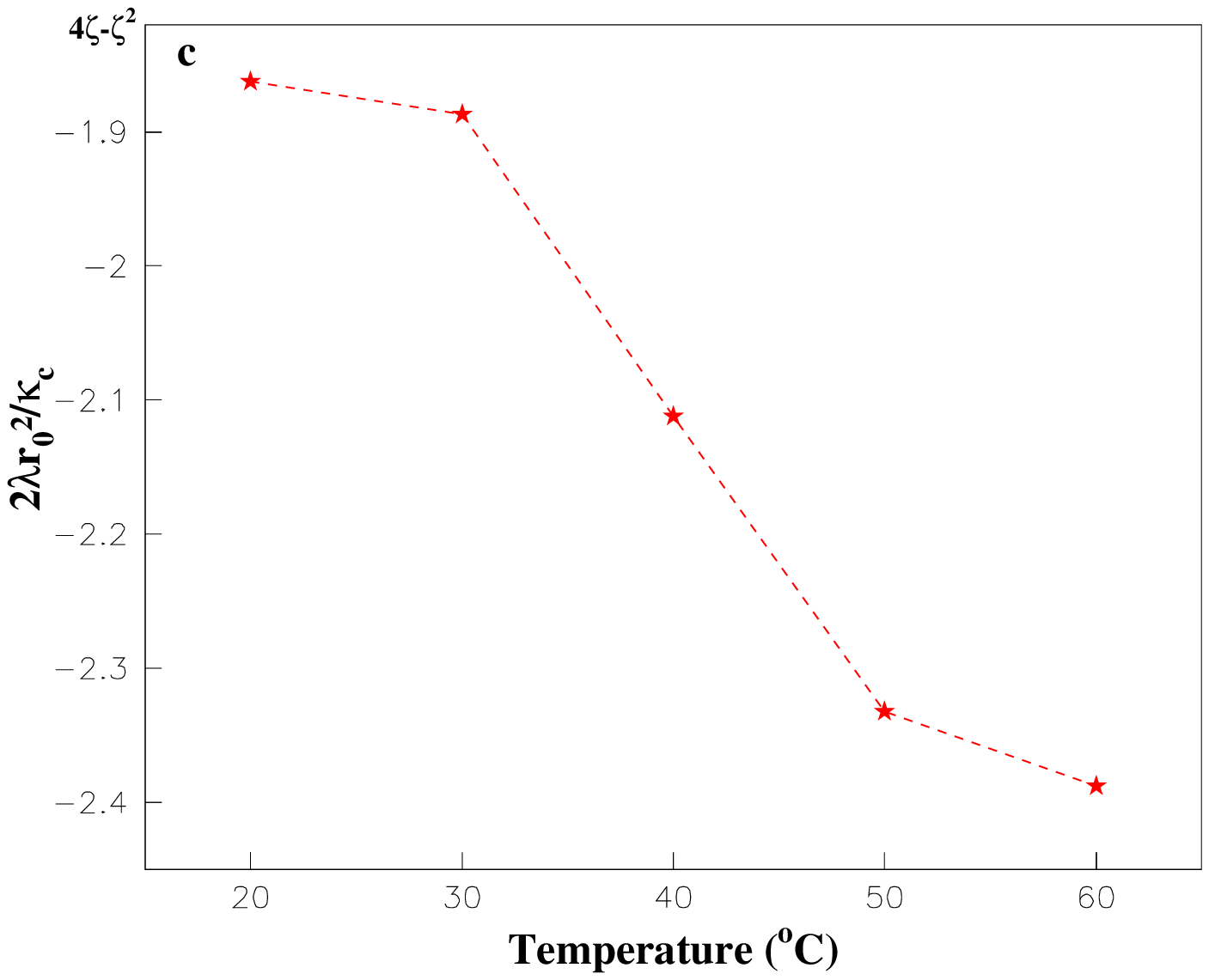}
\caption{\textbf{a.} The experimental data (solid lines) of birefringence at different temperature\cite{manyuhina07} are fitted by Eq.(\ref{eq:57})with $\Delta \chi>0$ (dash lines). The fitted parameters $\beta$ and $\eta$ are listed in table \ref{t:1}. \textbf{b.} The predicted value of birefringence with the parameters listed in table \ref{t:1} but $\Delta \chi<0$. \textbf{insert.} The deformation of the vesicle under different $B$ by Eq.(\ref{eq:50}) at 20$^{\rm o}$C. \textbf{c.} The influence of temperature on the $\kappa_{\rm c}$ can be estimated by Eq.(\ref{eq:570}). Results show that the influence between $30^{\rm o}$C and $50^{\rm o}$C is sensitive.}
\label{f1}
\end{figure}

\subsection{Case 1: light birefringence without the constraint of constant of surface area}\label{s4.1}

If there is not the constraint of constant surface area, that is $\delta \lambda=0$, Eq.(\ref{eq:18}) $=$ Eq.(\ref{eq:19}) will leads to:
\begin{eqnarray}
a_0&=&-\frac{qr_0B^2\sqrt{4\pi}}{\dfrac{2\lambda r_0^2}{\kappa_{\rm c}}+\zeta^2-4\zeta}\nonumber\\
&\equiv&\eta r_0B^2\sqrt{4\pi}\label{eq:48}\\
a_2&=&-\frac{2qr_0B^2\sqrt{\dfrac{4\pi}{5}}}{\dfrac{2\lambda r_0^2}{\kappa_{\rm c}}+\zeta^2-4\zeta+12}\nonumber\\
&\equiv&-\beta r_0B^2\sqrt{\frac{4\pi}{5}}\label{eq:49}
\end{eqnarray}
where $q\equiv t\Delta\chi r_{\rm 0}^2/(3\kappa_{\rm c}\mu)$, $\zeta\equiv c_0r_0$, $\eta\equiv q/(4\zeta-\zeta^2-2\lambda r_0^2/\kappa_{\rm c})$, and $\beta\equiv 2q/(2\lambda r_0^2/\kappa_{\rm c}+\zeta^2-4\zeta+12)$
\begin{eqnarray}
r_1(\theta)&=&|\vec{\boldsymbol{r}}_0+\psi\vec{\boldsymbol{n}}|\nonumber\\
&=&r_0+a_0Y_0+a_2Y_2\nonumber\\
&=&r_0\left[1+\left(\eta-\beta P_2(\cos\theta)\right)B^2\right]\label{eq:50}
\end{eqnarray}

Helfrich also suggested that the predicted deformation could be experimentally accessed through the field-induced birefringence of a suspension of identical vesicles, since the normalised birefringence\cite{helfrich732,manyuhina07,iwamoto13}
\begin{eqnarray}
\frac{\Delta n}{\Delta n_{\rm max}}=\frac{r(\pi/2)-r(0)}{R}\label{eq:55}
\end{eqnarray}
where $\Delta n=n_\shortparallel-n_\perp$, and $n_\perp$ is the refractive index that is always perpendicular to the optic axis ($z$), and $n_\shortparallel$ is the refractive index that is always parallel to the optic axis.

We have found the size of self-assembled vesicle to vary considerable with temperature. We therefore make the assumption that the deformation of the vesicle takes place without the constraint of constant surface area, that is $\delta \lambda=0$. The average radius of the ellipsoid, thus,
\begin{eqnarray}
R&=&\sqrt{\frac{A}{4\pi}}=\sqrt{\frac{1}{3}[2r_1(\pi/2)r_1(0)+r_1^2(\pi/2)]}\nonumber\\
&\approx&r_0\left[1+\eta B^2\right]\label{eq:56}
\end{eqnarray}
Thus, the normalised birefringence
\begin{eqnarray}
\frac{\Delta n}{\Delta n_{\rm max}}=\dfrac{1.5\beta B^2}{1+\eta B^2}\label{eq:57}
\end{eqnarray}
which can be used to fit the experimental data in Fig.\ref{f1}. Furthermore, the influence of temperature on the bend modulus can be estimated by
\begin{eqnarray}
\frac{2\lambda r_0^2}{\kappa_{\rm c}}=4\zeta-\zeta^2-\frac{12}{2\eta/\beta+1}\label{eq:570}
\end{eqnarray}
\begin{table}
\caption{Experimental data of birefringence at different temperature\cite{manyuhina07} in Fig.\ref{f1}a are fitted by the corresponding $\beta$ and $\eta$ with Eq.(\ref{eq:57}). The third low data is directly calculated by $12/(2\eta/\beta+1)$, which can be used to estimate the bend modulus $\kappa_{\rm c}$ as shown in Fig.\ref{f1}c. }
\begin{tabular}{l|l|l||l}
\hline
T($^\circ$C) & $\beta (10^{-2})$& $\eta (10^{-2})$&$4\zeta-\zeta^2-2\lambda r_0^2/\kappa_{\rm c}$\\
\hline
20&1.125&3.063&1.863\\
30&1.500&4.019&1.887\\
40&2.036&4.766&2.112\\
50&3.450&7.151&2.332\\
60&4.862&9.786&2.388\\
\hline
\end{tabular}\label{t:1}
\end{table}

\subsection{Case 2: stomatocyte with the constraint of constant of surface area}\label{s4.2}
For conservative surface area,
\begin{eqnarray}\label{eq.51}
\delta A&=&\oint-2H\psi r_0^2\sin\theta{\rm d}\theta{\rm d}\phi\nonumber\\
&=&\oint2r_0 (a_0{\rm Y}_0+a_2{\rm Y}_2)\sin\theta{\rm d}\theta{\rm d}\phi\nonumber\\
&=&4\sqrt{\pi}r_0a_0\nonumber\\
&=&0,
\end{eqnarray}
which leads $a_0=0$, and $\psi=a_2Y_2$. Eq.(\ref{eq:18}) $=$ Eq.(\ref{eq:19}) will leads to:
\begin{eqnarray}
\delta\lambda&=&\frac{t\Delta\chi B^2}{6\mu}\label{eq:52}\\
a_2&=&-\beta r_0B^2\sqrt{\frac{4\pi}{5}}\label{eq:53}
\end{eqnarray}
and
\begin{eqnarray}
r_2(\theta)&=&r_0+a_2Y_2\nonumber\\
&=&r_0[1-\beta P_2(\cos\theta)B^2]\label{eq:54}
\end{eqnarray}

\begin{figure}[htb]
\centering
\includegraphics[width=8cm]{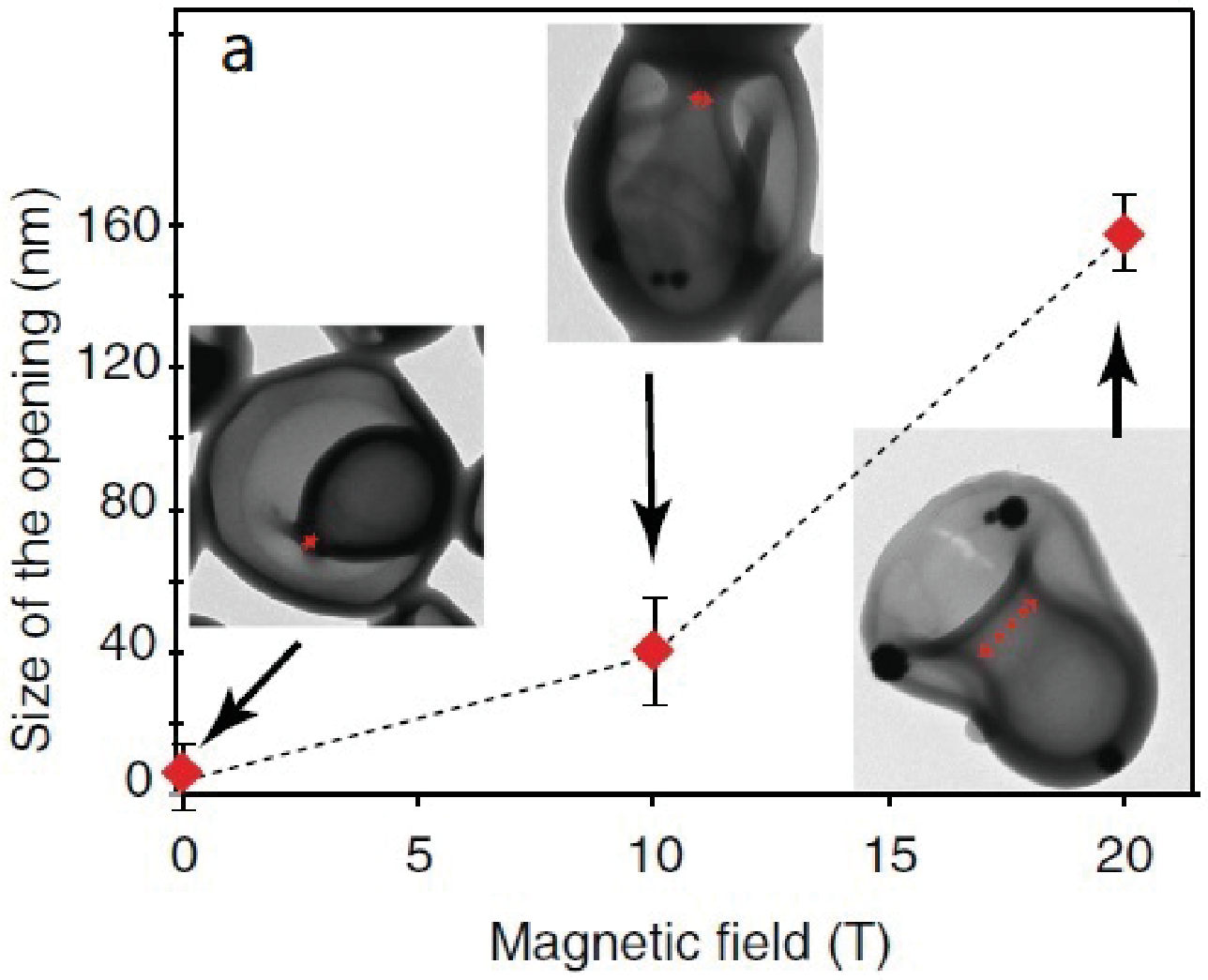}
\includegraphics[width=8cm]{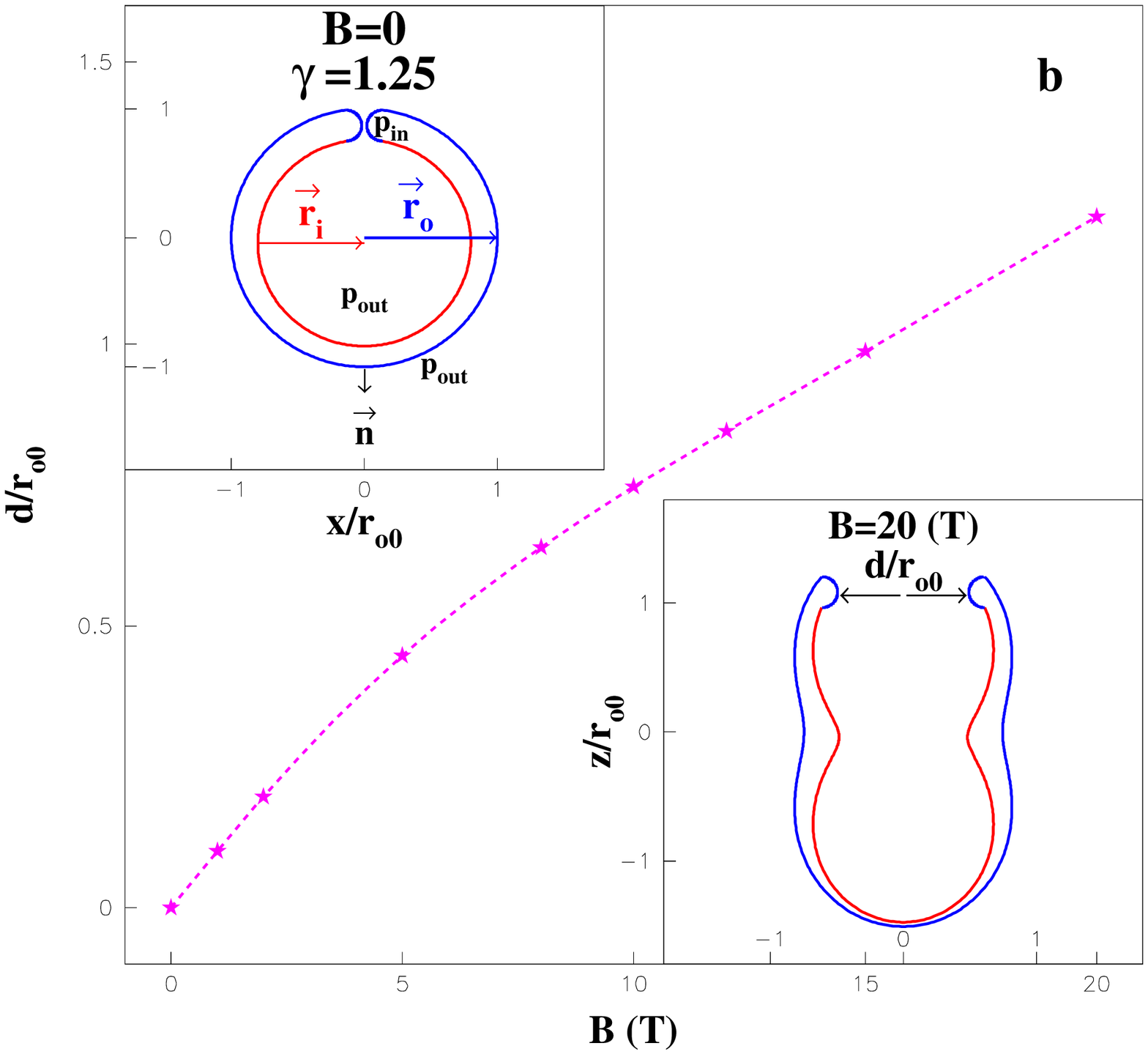}
\caption{\textbf{a.} The mouth of the artificial stomatocyte, which is assembled from diamagnetic amphiphilic block-copolymers with a highly anisotropic magnetic susceptibility, can be reversibly opened and closed by varying a external homogeneous magnetic fields\cite{Rhee14}.  \textbf{b.} The quantitative relation between the normalized open of the stomatocyte $d$ and $B$ can be calculated by Eqs (\ref{eq:63}) and (\ref{eq:64}). Here, $\zeta_{\rm o}=-1.5$, $t=26$ nm\cite{Rhee14}, $r_{\rm o0}=150$ nm\cite{Rhee14}, $r_{\rm i0}=120$ nm, $\kappa_{\rm c}=2.6\times10^{-21}$ J\cite{manyuhina07}, $\Delta\chi\approx-2.0\times10^{-7}$\cite{Rhee14,sutter69}, and $\mu_{\rm water}\approx 1.26\times10^{-6}$ N$\cdot$A$^{-2}$. \textbf{insert.} The deformation of the artificial stomatocyte under different magnetic induction $B$ by Eq.(\ref{eq:62}). For comfortable looking, we have moved the inside vesicle down and add the edge to connect two vesicles.}
\label{f2}
\end{figure}

A stomatocyte can be regarded as a system with two inner tangential vesicles. The shape of the cross section of the stomatocyte is crescent as shown in Fig.\ref{f2} b(\textbf{top inserted}). The outside spherical vesicle is convex, and its curvature $H=-1/r_{\rm o}$, thus $r_{{\rm o}0}$ and $r_{{\rm o}}$ correspond to Eq.(\ref{eq:11}) and Eq.(\ref{eq:54}) respectively. The inside one, however, is concave, and its curvature $H=1/r_{\rm i}$  because $\vec{\boldsymbol{r}}_{\rm i}/r_{\rm i}=-\vec{\boldsymbol{n}}$. Thus, we can simplify its calculation by replacing $r$ with $-r_{\rm i}$.

We assume that the small spherical vesicle (concave red solid: $r_{\rm i0}$) is internally tangent to the big one (convex blue solid: $r_{\rm o0}$) at $B=0$ as shown as in Fig.\ref{f1} b(\textbf{top inserted}). Then, we can substitute $r_{\rm o0}$ and $-r_{\rm i0}$ to the $r_0$ in Eq.(\ref{eq:11}) respectively\cite{iwamoto04},
\begin{eqnarray}
\Delta pr^2_{\rm o0}+(2\lambda+\kappa_{\rm c}c_0^2)r_{\rm o0}-2\kappa_{\rm c}c_0&=&0,\label{eq:58}\\
\Delta pr^2_{\rm i0}-(2\lambda+\kappa_{\rm c}c_0^2)r_{\rm i0}-2\kappa_{\rm c}c_0&=&0,\label{eq:59}
\end{eqnarray}
and get very important relations:
\begin{eqnarray}
\Delta p&=&\frac{2\kappa_{\rm c}c_0^3\gamma}{\zeta_{\rm o}^2},\label{eq.60}\\
\frac{2\lambda}{\kappa_{\rm c}}&=&-c_0^2\left[1+\frac{2(\gamma-1)}{\zeta_{\rm o}}\right],\label{eq:61}
\end{eqnarray}
where $\zeta_{\rm o}\equiv c_0r_{\rm o0}<0$ and $\gamma\equiv r_{\rm o0}/r_{\rm i0}>1$.

The two new deformation equations due to external magnetic field can be derived from Eq.(\ref{eq:50}) respectively
\begin{eqnarray}
\left\{\begin{array}{lcl}
r_{\rm o}(\theta_{\rm o})&=&r_{\rm o0}\left[1-\dfrac{q{\rm P}_2(\cos\theta_{\rm o})B^2}{6-(\gamma+1)\zeta_{\rm o}}\right]\\
r_{\rm i}(\theta_{\rm i})&=&r_{\rm i0}\left[1-\dfrac{q{\rm P}_2(\cos\theta_{\rm i})B^2}{6\gamma^2+(\gamma+1)\zeta_{\rm o}}\right]
\end{array}\right.\label{eq:62}
\end{eqnarray}
The deformation of the stomatocyte with different $B$ is calculated with Eq.(\ref{eq:62}) as shown in Fig.\ref{f2} b(\textbf{bottom inserted})

The size of mouth of the artificial stomatocyte, that is, the diameter $d$, can be determined by the equations:
\begin{eqnarray}
d\equiv2r_{\rm o}\sin(\theta_{\rm o})&=&2r_{\rm i}\sin(\theta_{\rm i})\label{eq:63}\\
r_{\rm o}\cos(\theta_{\rm o})&=&r_{\rm i}\cos(\theta_{\rm i})+r_{\rm o0}-r_{\rm i0}.\label{eq:64}
\end{eqnarray}
The relation between the size of mouth of the artificial stomatocyte ($d$) and the external homogeneous magnetic field ($B$) is calculated with Eqs. (\ref{eq:63}) and (\ref{eq:64}), and shown in Fig.\ref{f2} b.

\section{discusstion}
The perturbation $\psi$ for two cases have been derived into Eq.(\ref{eq:50}) and Eq.(\ref{eq:54}) respectively, which can be used to describe the small deformation of a vesicle due to the interaction of magnetic field. This deformation can be measured by birefringence as shown in Section \ref{s4.1}, and explain the reversibly open and closing of stomatocyte as in Section \ref{s4.2}. The fitted results shows that the influence of temperature between $30^{\rm o}$C and $50^{\rm o}$C on bend modulus is sensitive as shown in Fig.\ref{f1}c. Our calculation presupposes that only water can permeates transmembrane, and the changes of $\Delta p$ has been neglected. It should be pointed that this method is not situated the large deformation as shown in Fig.\ref{f1}.

\section{Acknowledgement}
The authors acknowledge the financial support by National
Natural Science Foundation of China (Grants No.11675180,
No. 11574329, and No. 11774358), Key Research Program
of Frontier Sciences of CAS (Grant No. Y7Y1472Y61), the
CAS Biophysics Interdisciplinary Innovation Team Project
(Grant No.2060299), the CAS Strategic Priority Research
Program (Grant No. XDA17010504), and the Joint NSFC-ISF
Research Program (Grant No. 51561145002).

\section*{Appendix: spherical vesicle differential geometry}
For spherical vesicle, we can use the spherical coordinates.
\begin{eqnarray}
\vec{\boldsymbol{r}}&\equiv& r_0\left(\cos\phi \sin\theta, \sin\phi \sin\theta, \cos\theta\right)\nonumber\\
\vec{\boldsymbol{\mathcal{H}}}&\equiv& \frac{B}{\mu}\left(0,0,1\right)\nonumber
\end{eqnarray}
then:
\begin{eqnarray}
\vec{\boldsymbol{r}}_\phi&=&r_0\left(-\sin\phi \sin\theta, \cos\phi \sin\theta, 0\right)\nonumber\\
\vec{\boldsymbol{r}}_\theta&=&r_0\left(\cos\phi \cos\theta, \sin\phi \cos\theta, -\sin\theta\right)\nonumber
\end{eqnarray}
\begin{eqnarray}
\vec{\boldsymbol{\mathcal{H}}}\cdot\vec{\boldsymbol{n}}&=&\frac{B}{\mu}\cos\theta, \nonumber\\
\vec{\boldsymbol{\mathcal{H}}}\cdot\vec{\boldsymbol{r}}_\phi&=&0, \nonumber\\ \vec{\boldsymbol{\mathcal{H}}}\cdot\vec{\boldsymbol{r}}_\theta&=&-r_0\frac{B}{\mu}\sin\theta\nonumber
\end{eqnarray}
\begin{eqnarray}
g^{\phi\phi}&=&\frac{1}{r_0^2\sin^2\theta}, \nonumber\\
g^{\theta\theta}&=&\frac{1}{r_0^2},\nonumber\\
g^{\phi\theta}&=&g^{\theta\phi}=0\nonumber
\end{eqnarray}
\begin{eqnarray}
L^{\phi\phi}&=&-\frac{1}{r_0\sin^2\theta},\nonumber\\
L^{\theta\theta}&=&-\frac{1}{r_0}, \nonumber\\
L^{\phi\theta}&=&L^{\theta\phi}=0\nonumber
\end{eqnarray}
\begin{eqnarray}
H_0&=&-\frac{1}{r_0}, \nonumber\\
K_0&=&\frac{1}{r_0^2}\nonumber
\end{eqnarray}
\begin{eqnarray}
\nabla&=&g^{ij}\vec{\boldsymbol{r}}_{\rm i}\partial_j=\frac{1}{r_0^2\sin^2\theta}\vec{\boldsymbol{r}}_\phi\partial_\phi+\frac{1}{r_0^2}\vec{\boldsymbol{r}}_\theta\partial_\theta\nonumber\\
\nabla^\star&=&L^{ij}\vec{\boldsymbol{r}}_{\rm i}\partial_j=-\frac{1}{r_0\sin^2\theta}\vec{\boldsymbol{r}}_\phi\partial_\phi-\frac{1}{r_0}\vec{\boldsymbol{r}}_\theta\partial_\theta\nonumber
\end{eqnarray}
\begin{eqnarray}
\vec{\boldsymbol{\mathcal{H}}}\cdot\nabla&=&-\frac{B}{\mu}\frac{\sin\theta}{r_0}\partial_\theta, \nonumber\\ \vec{\boldsymbol{\mathcal{H}}}\cdot\nabla^\star&=&\frac{B}{\mu}\sin\theta\partial_\theta\nonumber\\
\vec{\boldsymbol{\mathcal{H}}}\cdot\nabla^\star&=&-r_0\vec{\boldsymbol{\mathcal{H}}}\cdot\nabla\nonumber
\end{eqnarray}
\begin{eqnarray}
g&=&r^4_0\sin^2\theta, \nonumber\\
L&=&r^2_0\sin^2\theta, \nonumber\\
\nabla^2 H_0&=&0\nonumber
\end{eqnarray}

\end{document}